\documentclass[pre,superscriptaddress,twocolumn]{revtex4}
\usepackage{epsfig}
\usepackage{latexsym}
\usepackage{amsmath}
\begin {document}
%\title {Prisoner's dilemma with learning and evolving players}
\title {Statistical mechanics approach to a reinforcement learning model with memory}
\author{Adam Lipowski}
\affiliation{Faculty of Physics, Adam Mickiewicz University,
61-614 Pozna\'{n}, Poland}
\author{Krzysztof Gontarek}
\affiliation{Faculty of Physics, Adam Mickiewicz University,
61-614 Pozna\'{n}, Poland}
%\author{Dorota Lipowska}
%\affiliation{Institute of Linguistics, Adam Mickiewicz University,
%61-874 Pozna\'{n}, Poland}
\author{Marcel Ausloos}
\affiliation{GRAPES, University of Li\`ege, B-4000 Li\`ege,
Belgium}
%%%%%%%%%%%%%%%%%%%%%%%%%%%%%%%%%%%%%%%%%%%%%%%%%%%%%%%%%%%%%%%%%%%%%%%%%%%%%
\pacs{} \keywords{}
\begin {abstract}
We introduce a two-player model of reinforcement learning with
memory. Past actions of an iterated game are stored in a memory
and used to determine player's next action. To examine the
behaviour of the model some approximate methods are used and
confronted against numerical simulations and exact master
equation. When the length of memory of players increases to
infinity the model undergoes an absorbing-state phase transition.
Performance of examined strategies is checked in the prisoner'
dilemma game. It turns out that it is advantageous to have a large
memory in symmetric games, but it is better to have a short memory
in asymmetric ones.
\end{abstract}
\maketitle
\section{Introduction}
Game theory plays an increasingly important role in many
disciplines such as sociology, economy, computer sciences or even
philosophy~\cite{FUDENBERG}. Providing a firm mathematical basis,
this theory stimulates development of quantitative methods to
study general aspects of conflicts, social dilemmas, or
cooperation. At the simplest level such situations can be
described in terms of a two-person game with two choices. In the
celebrated example of such a game, the Prisoner's Dilemma, these
choices are called cooperate (C) and defect (D). The single Nash
equilibrium, where both players defect, is not Pareto optimal and
in the iterated version of this game players might have some
incentives to cooperate. However, finding an efficient strategy
even for such a simple game is highly nontrivial albeit exciting
task, as evidenced by the popularity of Axelrod's
tournaments~\cite{AXELROD}. These tournaments had the
unquestionable winner - the strategy tit-for-tat. Playing in a
given round what an opponent played in the previous round, the
strategy tit-for-tat is a surprising match of effectiveness as
well as simplicity. Later on various strategies were examined:
deterministic, stochastic, or evolving in a way that mimic
biological evolution. It was also shown that some strategies
perform better than the strategy tit-for-tat, as an example one
can mention the strategy called \emph{win-stay
loose-shift}~\cite{NOWAK}. In an interesting class of some other
strategies previous actions are stored in the memory and used to
determine future actions. However, since the number of possible
previous actions increases exponentially fast with the length of
memory and a strategy has to encode the response for each of such
possibilities, the length of memory has to be very
short~\cite{GOLBECK}. Such a short memory cannot detect possible
longer-term patterns or trends in the actions of the opponent.

Actually, the problem of devising an efficient strategy that would
use the past experience to choose or avoid some actions is of much
wider applicability, and is known as reinforcement learning.
Intensive research in this field resulted in a number of
models~\cite{LASLIER}, but mathematical foundations and analytical
insight into their behaviour seems to be less developed. Much of
the theory of the reinforcement learning is based on the Markov
Decision Processes where it is assumed that the player environment
is stationary~\cite{HOWARD}. Extension of this essentially
single-player problem to the case of two or more players is more
difficult but some attempts have been already made~\cite{LITTMAN}.
Urn models~\cite{BEGGS}  and various buyers-sellers
models~\cite{WALDECK} were also examined in the context od
reinforcement learning.

In most of the reinforcement learning models~\cite{EREV,BUSH} past
experience is memorized only as an accumulated payoff. Although
this is an important ingredient, storing the entire sequence of
past actions can potentially be more useful in devising efficient
strategies. To get a preliminary insight into such an approach, in
the present paper we introduce a model of an iterated game between
two players. A player stores in its memory the past actions of an
opponent and uses this information to determine probability of its
next action. We formulate approximate methods to describe the
behaviour of our model and confront them against numerical
simulations and exact master equation. Let us notice that
numerical simulations are the main and  often the only tool in the
study of reinforcement learning models. The possibility to use
analytical and sometimes even exact approaches such as those used
in the present paper seems to be a rare exception. Our
calculations show that when the length of memory increases to
infinity, a transition between different regimes of our model
takes place, that is analogous to an absorbing-state phase
transition~\cite{ODOR}. Similar phase transitions might exist in
spatially extended, multi-agent systems~\cite{HAUERT}, however in
the introduced two-player model this transition  has a much
different nature, namely it takes place only in the space of
memory configurations.

\section{A reinforcement learning model with memory}
\begin{figure}
\vspace{2cm} \centerline{ \epsfxsize=9cm \epsfbox{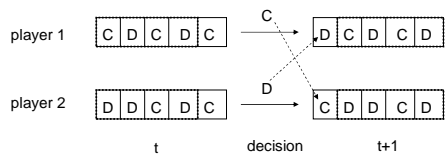} }
\vspace{-6cm} \caption{Memory change during a single round of a
game with two players with memories of length $l=5$.The first
player shifts all memory cells to the right (removing the
rightmost element) and puts the last decision (D) of the second
player at the left end. Analogous change takes place in the memory
of the second player} \label{player}
\end{figure}

In our model we consider a pair of players playing repeatedly a
game like e.g., the prisoner's dilemma. A player $i$ $(i=1,2)$ is
equipped with a memory of length $l_i$, where it sequentially
stores the last $l_i$ decisions made by its opponent. For
simplicity let us consider a
 game with two decisions that we denote as C and D. An example
 that illustrates a memory change in a single round of a game is
 shown in Fig.~\ref{player} (we will mostly examine the symmetrical case
where $l_1=l_2$, and the index $i$ denoting the player will be
thus dropped).

A player uses the information in its memory to evaluate the
opponent's behaviour and to calculate probabilities of making its
own decisions. Having in mind a possible application to the
prisoner's dilemma we make the eagerness to cooperate of a player
to be dependent on the frequency  of cooperation of its opponent.
More specifically, we assume that the probability $p_t$ for a
player to play C at the time $t$ is given by
\begin{equation}
p_t=1-a{\rm e}^{-bn_t/l}, \label{eq1}
\end{equation}
where $n_t$ is the number of $C$'s in player's memory at time $t$
while $a>0,b>0$ are some additional parameters. In principle $a$
can take any value such that $0 < a\leq 1$ but numerical
calculations presented below were made only for $a=1$ that left us
with only two control parameters, namely $b$ and $l$, that
determine the behaviour of the model. For $a=1$ the model has an
interesting absorbing state: provided that both players have
$n_t=0$ they both have $p_t=0$ and thus they will be forever
trapped in this (noncooperative) state. As we will see, this
feature in the limit $l\rightarrow\infty$ leads to a kind of phase
transition (already in the case of two players).

The content of the memory in principle might provide much more
valuable information on the opponent behaviour than
Eq.~\eqref{eq1} which is only one of the simplest possibilities.
As we already mentioned, our choice of the cooperation
probability\eqref{eq1} was motivated by the Prisoner's Dilemma but
of course for other games different expressions might be more
suitable. Moreover, more sophisticated expressions, for example
based on some trends in the distribution of C's, might lead to
more efficient strategies but such a possibility is not explored
in the present paper.

Let us also notice, that in our approach the memory of a player
stores the sequence of past actions of length $l$ (and that
information is used to calculate the probability of cooperation).
We do not store the response to each possible past sequence of
actions (as e.g., in~\cite{GOLBECK}) and that is why memory
requirements in our model increase only linearly with $l$ and not
exponentially.
%%%%%%%%%%%%%%%%%%%%%%%%%%%%%%%%%%%%%%%%%%%%%%%%%%%%%%%%%%
%%%%%%%%%%%%%%%%%%%%%%%%%%%%%%%%%%%%%%%%%%%%%%%%%%%%%%%%%%%%%%%%%%%%
\subsection{Mean-value approximation}
Despite a simple formulation the analysis of the model is not
entirely straightforward. This is mainly because the probability
$p_t$ is actually a random variable that depends on the
dynamically determined content of a player's memory. However, some
simple arguments can be used to determine the evolution of $p_t$
at least for large $l$. Indeed, in such a case one might expect
that fluctuations of $n_t/l$ are negligible and it might be
replaced in Eq.~\eqref{eq1} with its mean value. Since at time $t$
the coefficient $n_t$ of player (1) equals to the number of $C$'s
made by its opponent (2) during $l$ previous steps we obtain the
following expression for its mean value
\begin{equation}
\langle n_t^{(1)}\rangle=\sum_{k=1}^{l} p_{t-k}^{(2)}, \label{eq2}
\end{equation}
where the upper indices denote the players. Under such an
assumption we obtain that the evolution of probabilities
$p_{t}^{(1,2)}$ is given by the following equations
\begin{equation}
p_{t}^{(1,2)}=1-\exp \left({\frac{-b}{l}\sum_{k=1}^{l}
p_{t-k}^{(2,1)}}\right) \ \ t=l+1, l+2,\ldots. \label{eq3}
\end{equation}
In Eq.~\eqref{eq3} we assume that both players are characterized
by the same values of $b$ and $l$, but generalization to the case
where these parameters are different is straightforward. To
iterate Eq.~\eqref{eq3} we have to specify $2l$ initial values.
For the symmetric choice
\begin{equation}
p_{t}^{(1)}=p_{t}^{(2)} ,\ \ \ \ t=1,2\ldots,l, \label{symmetric}
\end{equation}
we obtain symmetric solutions (i.e., with Eq~\eqref{symmetric}
being satisfied for any $t$). In such a case the upper indices in
Eq.~\eqref{eq3} can be dropped.

\begin{figure}
\vspace{0cm} \centerline{ \epsfxsize=9cm \epsfbox{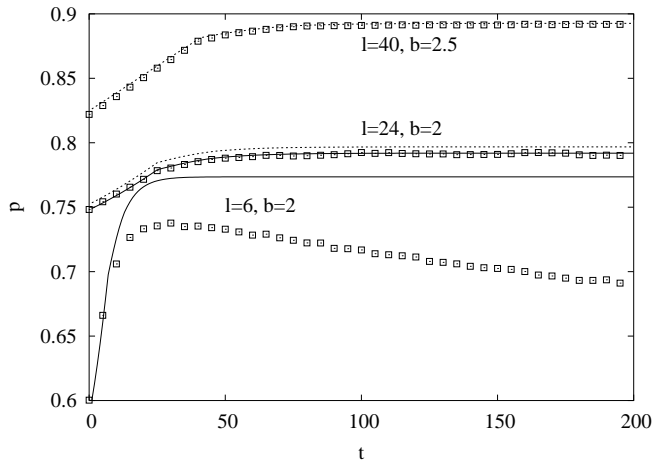} }
\caption{The cooperation probability $p$ as a function of time
$t$. The dashed lines correspond to the mean-value approximation
\eqref{eq3} while the continuous line shows the solution of
independent-decisions approximations~\eqref{ida-1a}. Simulation
data ($\Box$) are averages over $10^4$ independent runs. For
$l=24$ simulations and independent-decisions
approximation~\eqref{ida-1a} are in a very good agreement while
mean-value approximation~\eqref{eq3} slightly differs. For $l=40$
calculations using~\eqref{ida-1a} are not feasible but for such a
large $l$ a satisfactory description is obtained using the
mean-value approximation~\eqref{eq3}. Calculations for $l=6$ shows
that independent-decisions approximation deviates from
simulations. Results of approx.~\eqref{eq3} are not presented but
in this case they differ even more from simulation data. The
decrease of $p$ as seen in the simulation data is due to the the
small probability of entering an absorbing state (no cooperation).
On the other hand, approximations~\eqref{eq3} as well
as~\eqref{ida-1a} predict that for $t\rightarrow\infty$ the
probability $p$ tends to a positive value. For $l=24$ and 40 as
initial conditions we took (symmetric case) $p_t=0.7,\ \
t=1,2,\ldots, l$ and for $l=6$ we used $p_t=0.5$. Initial
conditions in Monte Carlo simulations corresponded to these
values. } \label{mfa-mc}
\end{figure}

For large $l$ the mean-value approximation \eqref{eq3} is quite
accurate. Indeed, numerical calculations show that already for
$l=40$ this approximation is in very good agreement with Monte
Carlo simulations (Fig.~\ref{mfa-mc}). However, for smaller $l$ a
clear discrepancy can be seen.

%%%%%%%%%%%%%%%%%%%%%%%%%%%%%%%%%%%%%%%%%%%%%%%%%%%%%%%%%%
%%%%%%%%%%%%%%%%%%%%%%%%%%%%%%%%%%%%%%%%%%%%%%%%%%%%%%%%%%%%%%%%%%%%%%%

Provided that in the limit $t\rightarrow\infty$ the system reaches
a steady state ($p_t=p$), in the symmetric case we obtain
\begin{equation}
p=1-\exp(-bp). \label{eq4}
\end{equation}
Elementary analysis show that for $b\leq 1$ the only solution
of~\eqref{eq4} is $p=0$ and for $b>1$ there is also an additional
positive solution. Such a behaviour typically describes a phase
transition at the mean-field level, but further discussion of this
point will be presented at the end of this section.
\subsection{Independent-decisions approximation}
As we already mentioned, the mean-value approximation~\eqref{eq3}
neglects fluctuations of $n_t$ around its mean value. In this
subsection we try to take them into account. Let us notice that a
player with memory length $l$ can be in one of the $2^{l}$
configurations ($conf$). Provided that we can calculate
probability $p_{{conf}}$ of being in such a configuration (at time
$t$), we can write
\begin{equation}
p_t= \sum_{{conf}}
\left[1-\exp\left(-\frac{bn({conf})}{l}\right)\right]p_{{conf}},\label{ida-1}
\end{equation}
where $n({conf})$ is the number of $C$'s in a given configuration
$conf$ and the summation is over all $2^{l}$ configurations;
indices of players are temporarily omitted. But for a given
configuration we know its sequence of C's and D's and thus its
history. For example, if at time $t$ a memory of a player (with
$l=3$) contains CDD it means that at time $t-1$ its opponent
played $C$ and at time $t-2$ and $t-3$ played D (we use the
convention that most recent elements are on the left side).
Assuming that such actions are independent, in the above example
the probability of the occurrence of this sequence might be
written as $p_t(1-p_{t-1})(1-p_{t-2})$. Writing $p_{conf}$ in such
a product form for arbitrary $l$, Eq.~\eqref{ida-1} can be written
as
\begin{equation}
p_t= \sum_{\{E_{k}\}}
\left[1-\exp\left(-\frac{bn(\{E_{k}\})}{l}\right)\right]\prod_{k=1}^{l}f_{t-k}(E_k),\label{ida-1a}
\end{equation}
where the summation in Eq.~\eqref{ida-1a} is over all $2^{l}$
configurations (sequences) $\{E_{k}\}$ where $E_k=$ C or D and
$k=1,\ldots,l$. Moreover, $n(\{E_{k}\})$ equals the number of C's
in a given sequence and
\begin{equation}
f_{t-k}(E_k)=\left\{
\begin{array}{ll}
p_{t-k} & {\rm for}\ \ E_k=C\\
1-p_{t-k} & {\rm for}\ \ E_k=D
\end{array}
\right. \label{ftk}
\end{equation}
For $l=2$, Eq.~\eqref{ida-1a} can be written as
{\setlength\arraycolsep{1pt}
\begin{eqnarray}
p_t^{(1,2)} & = &
p_{t-1}^{(2,1)}p_{t-2}^{(2,1)}r_2+p_{t-1}^{(2,1)}(1-p_{t-2}^{(2,1)})r_1+\nonumber\\
& & +
(1-p_{t-1}^{(2,1)})p_{t-2}^{(2,1)}r_1+\nonumber\\
& & + (1-p_{t-1}^{(2,1)})(1-p_{t-2}^{(2,1)})r_0, \label{ida-2}
\end{eqnarray}}
where $r_k=1-\exp{(-bk/2)}$.

The number of terms in the sum of~Eq.~\eqref{ida-1a} increases
exponentially with $l$, but numerically one can handle
calculations up to $l=24\sim28$. Solution of Eq.~\eqref{ida-1a} is
in much better agreement with
 simulations than the mean-value approximation\eqref{eq3}.
For example for $l=24$ and $b=2$ it essentially overlaps with
simulations, while~\eqref{eq3} clearly differs (Fig.\ref{mfa-mc}).

Despite an excellent agreement seen in this case, the
scheme~\eqref{ida-1a} is not exact. As we already mentioned, this
is because the product form of the probability $p_{conf}$ is based
on the assumption that decisions at time $t-1,t-2,\ldots, t-l$ are
independent, while in fact they are not. For smaller values of $l$
the (increasing in time) difference with simulation data might be
quite large (Fig.\ref{mfa-mc}).
%%%%%%%%%%%%%%%%%%%%%%%%%%%%%%%%%%%%%%%%%%%%%%%%%%%%%%%%%%
\begin{figure}
\vspace{0cm} \centerline{ \epsfxsize=9cm \epsfbox{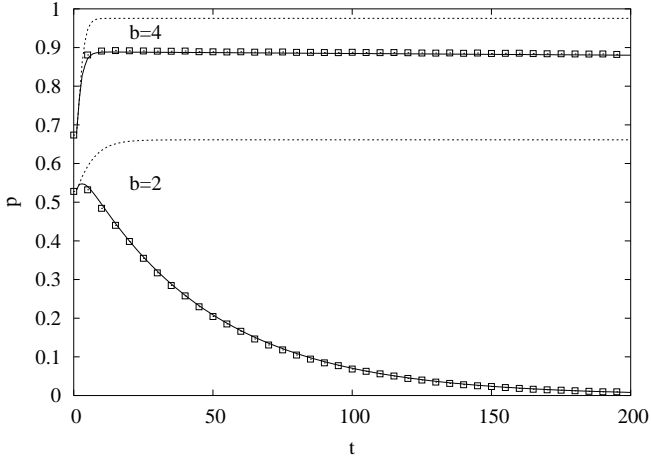} }
\caption{The cooperation probability as a function of time $t$ for
two players with $l=2$. Exact master equation
solution~\eqref{exact-ev-eq}-\eqref{probt1} (solid line) is in
perfect agreement with simulations ($\Box$) and deviates from the
independent-decisions approximation~\eqref{ida-2} (dotted line).}
\label{exact}
\end{figure}
%%%%%%%%%%%%%%%%%%%%%%%%%%%%%%%%%%%%%%%%%%%%%%%%%%%%%%%%%%%%%%%
\subsection{Master equation}
In this subsection we present the exact master equation of this
system. This equation directly follows from the stochastic rules
of the model and describes the evolution of probabilities of the
system being in a given state. Let us notice that a state of the
system is given by specifying the memory content of both agents.
In the following we present the explicit form of this equation
only in the case $l_{\rm}=2$, but an extension to larger $l$ is
straightforward but tedious. We denote the occupation probability
of being at time $t$ in the state where the first player has in
its memory the values E, F and the second one has G and H as
$p_t^{EF,GH}$. Assuming that the parameters $b$ and $l$ are the
same for both players and that symmetric initial conditions are
used
\begin{equation}
p_{t}^{EF,GH}=p_{t}^{GH,EF} ,\ \ \ \ t=0 \label{ex-symmetric}
\end{equation}
enables us to reduce the number of equations  from 16 to 10. The
resulting equations preserve the symmetry~\eqref{ex-symmetric} for
any $t$ and are the same for each of the players. The master
equation of our model for $t=1,2,\ldots$ takes the following form
\begin{eqnarray}
p_t^{\rm CC,CC}  & = & p_{t-1}^{\rm CC,CC}r_2^2+2p_{t-1}^{\rm
CC,CD}r_2r_1+{}\nonumber\\
& & {}+p_{t-1}^{\rm CD,CD}r_1^2\nonumber\\
 p_t^{\rm CC,CD}  & = & p_{t-1}^{\rm CC,CC}r_2r_1{}+p_{t-1}^{\rm CD,DC}r_1^2\nonumber\\
 p_t^{\rm CC,DC}  & = & p_{t-1}^{\rm CC,CC}r_2(1-r_2)+p_{t-1}^{\rm
CD,CD}r_1(1-r_1)+\nonumber\\
& & +p_{t-1}^{\rm CC,CD}(r_1+r_2 -2r_1r_2)\nonumber\\
p_t^{\rm CC,DD}  & = & p_{t-1}^{\rm CC,DC}r_1(1-r_2)+p_{t-1}^{\rm CD,DC}r_1(1-r_1)\nonumber\\
p_t^{\rm CD,DC}  & = & p_{t-1}^{\rm CC,DC}r_2(1-r_1)+p_{t-1}^{\rm
CC,DD}r_2+{}\nonumber\\
& & {}+ p_{t-1}^{\rm CD,DC}r_1(1-r_1)+p_{t-1}^{\rm CD,DD}r_1\nonumber\\
p_t^{\rm DC,DC}  & = & p_{t-1}^{\rm CC,CC}(1-r_2)^2+p_{t-1}^{\rm
CC,CD}(1-r_2)(1-r_1)+{}\nonumber\\
& & {}+ p_{t-1}^{\rm CC,CD}(1-r_2)(1-r_1)+p_{t-1}^{\rm CD,CD}(1-r_1)^2\nonumber\\
p_t^{\rm CD,CD}  & = & p_{t-1}^{\rm DC,DC}r_1^2\nonumber\\
p_t^{\rm DC,DD}  & = & p_{t-1}^{\rm CD,DD}(1-r_1)+p_{t-1}^{\rm
CC,DD}(1-r_2)+{}\nonumber\\
& & {}+ p_{t-1}^{\rm CD,DC}(1-r_1)^2+p_{t-1}^{\rm CC,DC}(1-r_1)(1-r_2)\nonumber\\
p_t^{\rm CD,DD}  & = & p_{t-1}^{\rm DC,DD}r_1+p_{t-1}^{\rm DC,DC}r_1(1-r_1)\nonumber\\
p_t^{\rm DD,DD}  & = & p_{t-1}^{\rm DD,DD}+2p_{t-1}^{\rm
DC,DD}(1-r_1)+{}\nonumber\\
& & {}+ p_{t-1}^{\rm DC,DC}(1-r_1)^2. \label{exact-ev-eq}
\end{eqnarray}
Iterating~Eq.~\eqref{exact-ev-eq} one can calculate all occupation
probabilities $p_t^{EF,GH}$. The result can be used to obtain the
probability of cooperating at time $t$
\begin{eqnarray}
p_{t} & = & p_{t+1}^{\rm CC,CC}+p_{t+1}^{\rm CC,DC}+2p_{t+1}^{\rm
CC,CD}+p_{t+1}^{\rm CC,DD}+\nonumber\\
& & +p_{t+1}^{\rm CD,CD}+p_{t+1}^{\rm CD,DC}+p_{t+1}^{\rm CD,DD}.
\label{probt1}
\end{eqnarray}
For $b=2$ and 4 the numerical results are presented in
Fig.~\ref{exact}. One can see that they are in perfect agreement
with simulations. Let us notice that for $b=2$ after a small
initial increase, the cooperation probability $p_t$ decreases in
time. This is an expected feature and is caused by the existence
of the absorbing state DD,DD. Of course, the
equations~\eqref{exact-ev-eq} reflect this fact: the probability
$p_{t-1}^{\rm DD,DD}$ enters only the last equation, namely that
describing the evolution of $p_t^{\rm DD,DD}$ (in other words,
none of the states can be reached from this state). Although on a
larger time scale $p_t$ would decrease also for $b=4$, on the
examined time scale it seems to saturate at a positive value.
Solutions (i.e., $p_t$) obtained  from the independent-decisions
approximation as well as mean-value approximation saturates at
some positive values in the limit $t\rightarrow\infty$ and thus
approximately correspond to such quasi-stationary states.

%%%%%%%%%%%%%%%%%%%%%%%%%%%%%%%%%%%%%%%%%%%%%%%%%%%%%%%%%%
\begin{figure}
\vspace{0cm} \centerline{ \epsfxsize=9cm \epsfbox{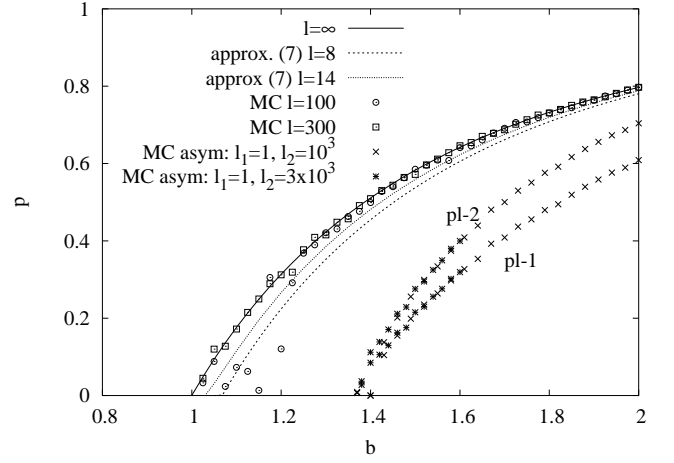} }
\caption{The steady-state cooperation probability $p$ as a
function of $b$. The independent-decision
approximation~\eqref{ida-1a} for increasing $l$ converges to the
mean-value approximation~\eqref{eq4} that in the limit $l=\infty$
presumably becomes exact. In the asymmetric case the cooperation
probability of each player is different. The first player (pl-1)
has the memory length $l_1=1$ and the second player (pl-2) has
$l_2=10^3$ or $3\cdot 10^3$.} \label{phasetr}
\end{figure}
%%%%%%%%%%%%%%%%%%%%%%%%%%%%%%%%%%%%%%%%%%%%%%%%%%%%%%%%%%%%%%%

The (quasi-)stationary behaviour of the model is presented in
Fig.~\ref{phasetr}. Provided that $b$ is large enough the players
remain in the cooperative phase; otherwise they enter the
absorbing (noncooperative) state. However, for finite memory
length $l$ the cooperative state is only a transient state, and
after a sufficiently large time an absorbing state will be
reached. Thus, strictly speaking, a phase transition between
cooperative and noncooperative regimes takes place only in the
limit $l\rightarrow\infty$. In this limit the mean-value
approximation~\eqref{eq4} correctly describes the behaviour of the
model. Simulations agree with~\eqref{eq4}, but to obtain good
agreement for $b$ close to the transition point value $b=1$, the
length of memory $l$ should be large.

We have also examined the nonsymmetric (with respect to the memory
length) case, where the first player has the memory of finite
length $l_1$ and the length of the memory of the second player
$l_2$ diverges. Simulations for $l_1=1$ and $l_2=10^3$ and $3\cdot
10^3$ show that in this case there is also a phase transition
(Fig.~\ref{phasetr}) but at a larger value of $b$ than in the
symmetric case (apparently, fluctuations due to the short memory
of the first player ease the approach of an absorbing state).
Results for larger values of $l_1$ (not presented) show that this
transition approaches the phase transition in the symmetric case.

The phase transition that is shown in Fig.~\ref{phasetr} is an
example of an absorbing-state phase transition with cooperative
and noncooperative phases corresponding to active and absorbing
phases, respectively~\cite{ODOR}. Such transitions appear also for
some models of Prisoner's Dilemma (or other games) in spatially
extended systems~\cite{HAUERT}, i.e., the phase transition appears
in the limit when the number of players increases to infinity. In
the present model the nature of this transition is much different:
the number of players remains finite (and equal to two) but the
length of memory diverges.
\section{Prisoner's Dilemma}
\begin{figure}
\vspace{1cm} \centerline{ \epsfxsize=9cm \epsfbox{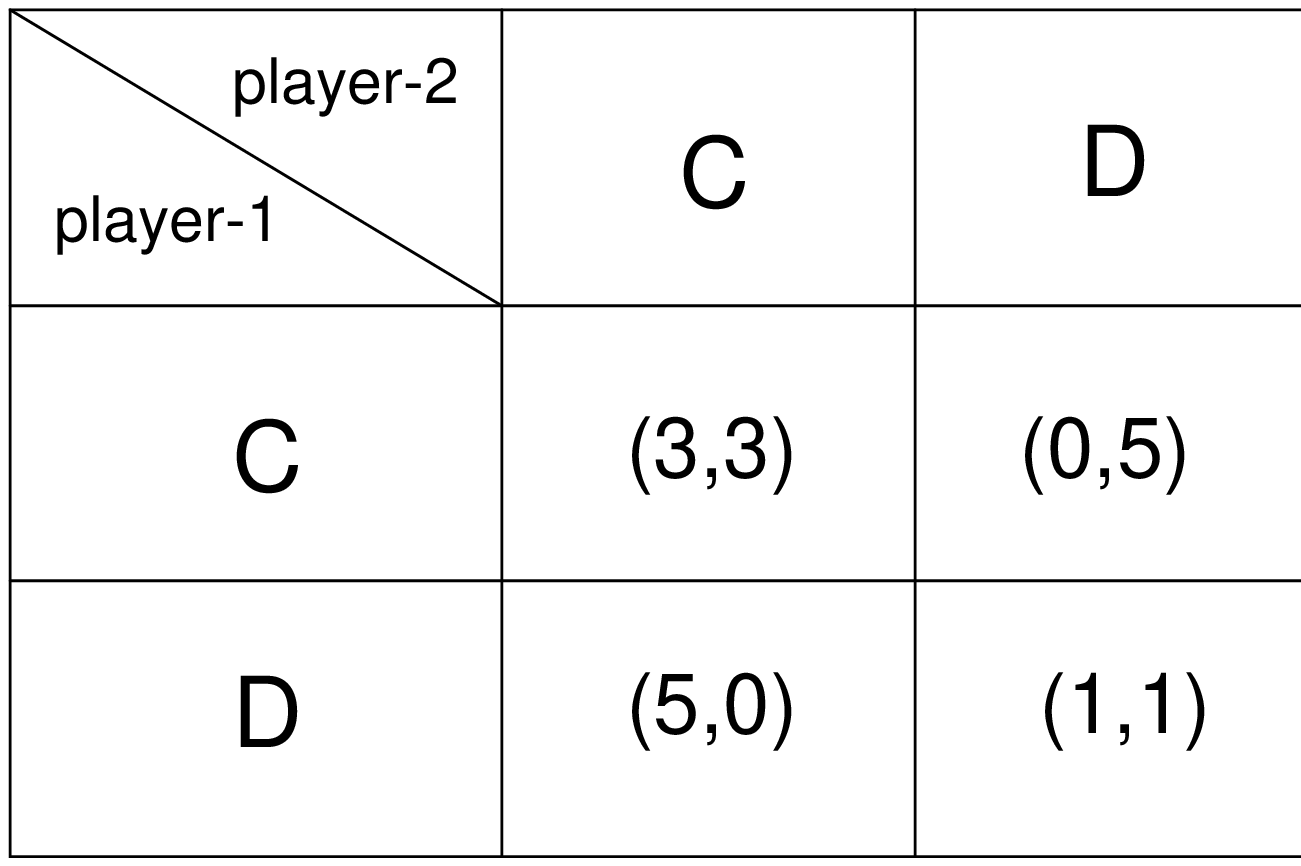} }
\vspace{-4cm} \caption{The payoff matrix of the prisoner's dilemma
game used in the calculations presented in
Figs.~\ref{paysymfig}-\ref{payasymfig}. The first and the second
number of a pair in a given cell denotes payoff of the first and
second player, respectively.} \label{payoff-matrix}
\end{figure}

\begin{figure}
\vspace{0cm} \centerline{ \epsfxsize=9cm \epsfbox{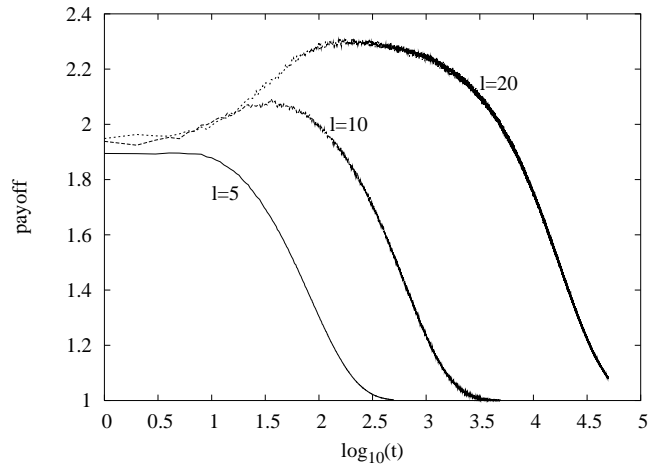} }
\vspace{-0cm} \caption{The time evolution of the average payoff in
the symmetric case ($l_1=l_2=l$) for $b=1.5$ and several values of
$l$. Results are averages over $10^5$ independent runs. As an
initial state each player at each cell of its memories has $C$ or
$D$ with probabilities $0.3$ and 0.7, respectively.}
\label{paysymfig}
\end{figure}

\begin{figure}
\vspace{0cm} \centerline{ \epsfxsize=9cm \epsfbox{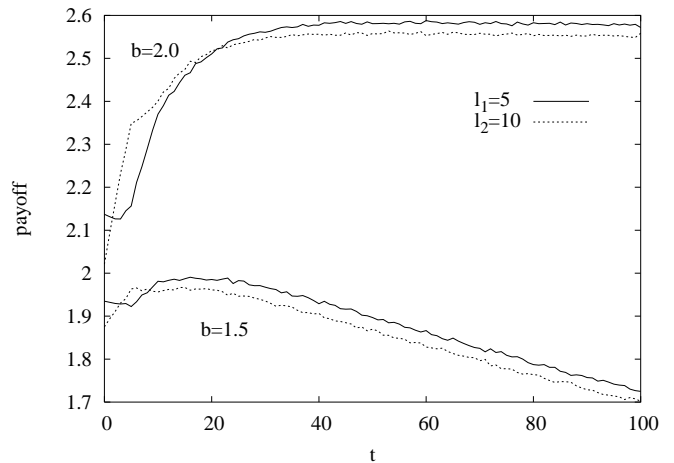} }
\vspace{-0cm} \caption{The time evolution of the average payoff in
the asymmetric case $l_1=5$ and $l_2=10$. Results are averages
over $10^5$ independent runs. As an initial state each player at
each cell of its memories has $C$ or $D$ with probabilities $0.3$
and 0.7, respectively.} \label{payasymfig}
\end{figure}

In this section we examine our players in an explicit example of
the Prisoner's Dilemma with the typically used payoff matrix that
is shown in Fig.~\ref{payoff-matrix}.
%%%%%%%%%%%%%%%%%%%%%%%%%%%%%%%%%%%%%%%%%%
%%%%%%%%%%%%%%%%%%%%%%%%%%%%%%%%%%%%%%
Results of the calculations of the time dependence of the average
payoff are presented in Figs.~\ref{paysymfig}-\ref{payasymfig}.
Simulations in the symmetric case (Fig.~\ref{paysymfig}) show that
the larger the memory length $l$, the larger the payoff. In the
asymmetric case (Fig~.~\ref{payasymfig}) the shorter-memory player
for large $t$ has larger payoff, but initially it might have the
smaller payoff than the longer-memory player. In simulations shown
in Figs.~\ref{paysymfig}-\ref{payasymfig} the memory length was
rather short and the model relatively quickly enters the absorbing
(noncooperative) state. That is why the average payoff converges
asymptotically to unity. Although this is not shown, such a
behaviour was seen also in the asymmetric case, but on a larger
time scale than that presented in Fig.~\ref{payasymfig}.
%%%%%%%%%%%%%%%%%%%%%%%%%%%%%%%%%%%%%%%%%%
%%%%%%%%%%%%%%%%%%%%%%%%%%%%%%%%%%%%%%
%%%%%%%%%%%%%%%%%%%%%%%%%%%%%%%%%%%%%%%%%%
%%%%%%%%%%%%%%%%%%%%%%%%%%%%%%%%%%%%%%

Using solely the results shown in
Figs.~\ref{paysymfig}-\ref{payasymfig} it is difficult to predict
what are the parameters ($l$, $b$) of the best (i.e., accumulating
the largest payoff) player. This is because the performance of a
given player depends on the parameters of the opponent, number of
rounds or even the initial content of the memory. And already the
length of memory alone results in conflicting properties: it pays
off to have a large memory in symmetric games
(Fig.~\ref{paysymfig}), but it is better to have a short memory in
asymmetric ones (Fig.~\ref{payasymfig}). It would be thus
interesting to perform Axelrod's type tournament that would make
the evolutionary selection of the winner,  where the accumulated
payoff of each player would determine its fitness. Particularly
interesting might be to examine a spatially extended version of
such a tournament, where opponents of a given player would be only
its neighbouring sites. In such a tournament one can check for
example whether spatial effects modify the nature (i.e.,
universality class) of the absorbing-state phase transition. And
of course, it would be interesting to check whether in such an
ensemble of players the strategy tit-for-tat, that in our model is
obtained for $l=1$ and $b\rightarrow\infty$, will be again
invincible.

As a further extension one can consider playing multi-decision
games. In such a case an additional group structure might appear
and examination of the nature of cooperation becomes much more
subtle~\cite{PHAN}.
\section{Conclusions}
In the present paper we have introduced a reinforcement learning
model with memory and have analysed it using approximate methods,
numerical simulations and exact master equation. In the limit when
the length of memory becomes infinite the model has an
absorbing-state phase transition. The objective of the paper was
to develop general approaches (such as approximate descriptions or
master-equation analysis) to study such models, and that is why
rather a simple and motivated by the Prisoner Dilemma form
(\ref{eq1}) of the cooperation probability, was used. In some
particular games more complicated functions might prove more
efficient. One can also consider storing in player's memory some
additional information concerning, e.g., players own moves.
Perhaps analytical approaches, that we used in some simple
examples, can be adapted to such more complicated problems as
well.

We also suggested that it would be desirable to perform Axelrod's
type tournament for players with memory (as in our work), but in
addition equipped with some evolutionary abilities~\cite{MIEKISZ}.
Such a tournament would allow us to examine the coexistence of
learning and evolution that is an interesting subject on its own.
Better learning abilities might influence the survival and thus
direct the evolution via the so-called Baldwin
effect~\cite{BALDWIN}. Some connections between learning and
evolution were already examined also in the game-theory
setup~\cite{HINGSTON,SUZUKI}. For the present model a detailed
insight at least into learning processes is available and coupling
them with evolutionary processes might lead to some interesting
results in this field.

Finally, let us notice that decision making based on the content
of memory seems to be connected with the psychophysical relation
between response and stimulus. Early attempts to express such a
relation in mathematical terms lead to the so-called Weber-Fechner
law~\cite{FECHNER}. Despite some works that reproduce this type of
law~\cite{COPELLI}, further research, perhaps using models similar
to those described in the present paper, would be desirable.
%%%%%%%%%%%%%%%%%%%%%%%%%%%%%%%%%%%%%%%%%%%%%%%%%%%%%%%%%

Acknowledgments: We gratefully acknowledge access to the computing
facilities at Pozna\'n Supercomputing and Networking Center. A.L.
and M.A were supported by the bilateral agreement between
University of Li\`ege and Adam Mickiewicz University.
%%%%%%%%%%%%%%%%%%%%%%%%%%%%%%%%%%%%%%%%%%%%%%%%%%%%%%%%%%%%%%%%%%%%%%%%%%%%%%

%%%%%%%%%%%%%%%%%%%%%%%%%%%%%%%%%%%%%%%%%%%%%%%%%%%%%%%%%%%%%%%%%%%%%%%%%%%%%%%
%%%%%%%%%%%%%%%%%%%%%%%%%%%%%%%%%%%%%%%%%%%%%%%%%%%%%%%%%%%%%%%%%%%%%%%%%%%%%%%
\end {document}